\documentclass[prl,aps,twocolumn,showpacs,superscriptaddress]{revtex4}
\usepackage[dvips]{graphicx}
\usepackage{latexsym}
\usepackage{amsmath}
\usepackage{subfigure}
\usepackage[hypertex]{hyperref}
\usepackage{color}

\usepackage{amssymb}
\begin{document}

\title{
The quantum valley Hall effect
in proximity-induced superconducting graphene:
an experimental window for deconfined quantum criticality
}
\author{Pouyan Ghaemi}
\affiliation{Department of Physics, University of California at
Berkeley, Berkeley, CA 94720, USA}
\author{Shinsei Ryu}
\affiliation{Department of Physics, University of California at
Berkeley, Berkeley, CA 94720, USA}
\author{Dung-Hai Lee}
\affiliation{Department of Physics, University of California at
Berkeley, Berkeley, CA 94720, USA} \affiliation{Material Science
Division, Lawrence Berkeley National Laboratory, Berkeley, CA 94720,
USA}

\date{\today}

\begin{abstract}
We argue that by inducing superconductivity in graphene via the proximity effect, it is
possible to observes the ``quantum valley Hall effect''. In the presence of magnetic field,
supercurrent causes ``valley pseudospin'' to accumulate at the edges of the superconducting
strip in Fig.\ \ref{setup}. This, and the structure of the superconducting vortex core,
provide possibilities to experimentally observe aspects of the  ``deconfined quantum criticality''.
\end{abstract}

%\pacs{73.43.-f, 73.20.At, 74.25.Fy, 73.20.Fz, 03.65.Vf}

\maketitle

\begin{figure}[htp]
\begin{center}
\subfigure[]{
\includegraphics[width=0.2 \textwidth]{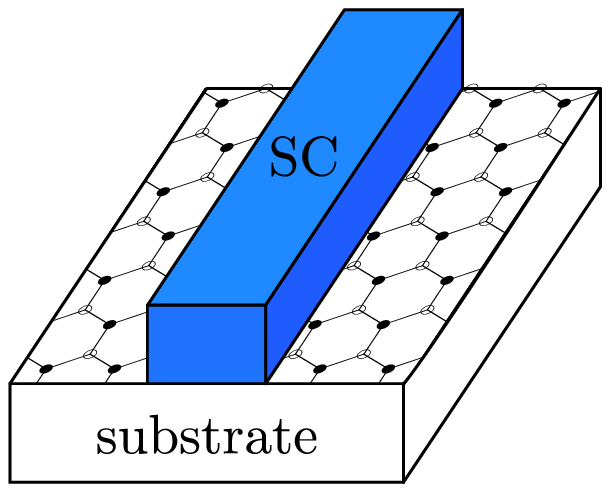}
} \subfigure[]{
\includegraphics[width=0.2 \textwidth]{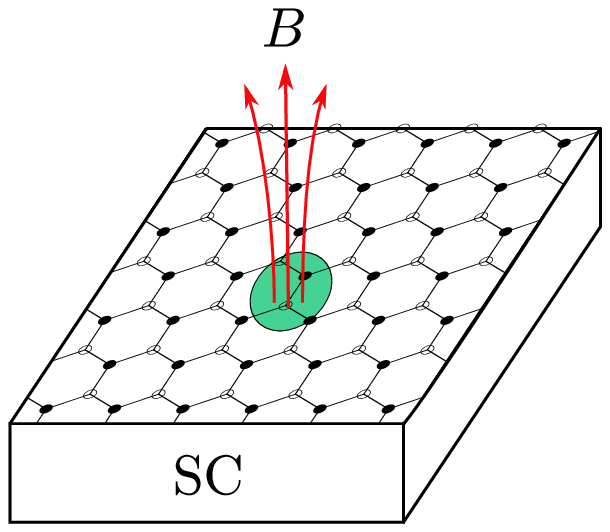}
}
 \caption{
Proximity induced superconductivity in graphene.
(a) Quantum valley Hall effect and
(b) superconducting vortex.
In (b), small staggered potential is
introduced near the core to detect accumulated valley spin.
\label{setup}
}
\end{center}
\end{figure}

Neutral graphene is a semimetal with Fermi points
($\mathbf{K}_{+}$ and $\mathbf{K}_{-}$) at the Brillouin zone corners\cite{graphene,graphenedirac}.
Vanishing density of states and low dimensionality tend to prevent graphene from developing order.
However, it has been demonstrated that through proximity effect by making a contact with SC electrodes,
graphene can carry dissipationless supercurrent\cite{graphenesc1,graphenesc2}
which is a promising observation for possibility of inducing superconductivity through proximity effect in graphene\cite{beenakker}
(Fig.\ \ref{setup}).

In this letter we show that if one views the valley index $\bf{K}_{\pm}$ as  a pseudospin degree of freedom (valley pseudospin), the proximity induced SC graphene will exhibit the ``quantum valley Hall effect'' (QVHE)
in the presence of magnetic field. The QVHE is a close cousin of the more familiar quantum spin Hall effect (QSHE) \cite{qshkm,qshe}. Here, supercurrent induces valley spin accumulation
at the edges of superconducting graphene.

This effect is closely related to the physics of the so-called deconfined quantum criticality\cite{dqc} where
continuous Landau-forbidden transition is made possible because
the defects of one phase carry the quantum number of the order parameter of the other phase.
Interestingly despite the fact that neither superconductivity nor ``Kekule distortion''\cite{Hou07} is realized (or even nearly being realized) in
graphene, the vortex core in the proximity induced SC graphene carry Kekule texture under suitable condition (see later). This ``duality'' between the superconducting and Kekule order is in sharp contrast with the 
usual competing order phenomena where nearly degenerate orders compete for realization in the ground state. As a consequence when one order is suppressed in the vortex core the other order emerges\cite{sachdev}.  If the SC order and the Kekule distortion are somehow stabilized in graphene, the above duality will make the continuous transition between the Kekule order and superconductivity possible.

The physical system we consider consists of a bulk $s$-wave superconductor coated by a single layer graphene
shown in Fig.\ \ref{setup}. Through the proximity effect, SC can be induced in graphene\cite{beenakker}.
 In addition since quasiparticles in graphene carry multiple quantum numbers, different types of paring could be induced. In the following we consider the simplest on-site $s$-wave pairing described by the following Hamiltonian
\begin{eqnarray}\label{schamilton}
&
\mathcal{H}
=
\displaystyle \int d\textbf{r}
\Psi^{\dag}
K
\Psi^{\ },
&
\nonumber \\%%%%%
&
K
=
\left(
\begin{array}{cc}
h & \Delta \Lambda \\
\Delta^* \Lambda^{-1} & -h^T\\
\end{array}
\right),
\quad
\Psi
=
\left(
\begin{array}{c}
\psi_{\uparrow}^{\ } \\
 \psi_{\downarrow}^{\dag T}
\end{array}
\right).&
\end{eqnarray}
Here
$\psi_s
=
\left(
\psi_{A+ s},
\psi_{B+ s},
\psi_{A- s},
\psi_{B- s}
\right)^T
$ is the four component fermionic field with spin $s$.
$\psi_{A/B, +/-, s}$
represents
the electron annihilation operator on sublattice $A/B$ associated with valley
$\mathbf{K}_{\pm}$
and spin $s$.
In the following, we use three sets of  Pauli matrices,
$\{\tau_{x,y,z,0}\}$,
$\{\sigma_{x,y,z,0}\}$,
and
$\{\rho_{x,y,z,0}\}$,
each acting on
sublattice ($A,B$),
valley ($+,-$)
and Nambu  ($\psi_{\uparrow}^{\ },  \psi_{\downarrow}^{\dag T}$) degrees of freedom,
respectively.
In Eq.\ (\ref{schamilton}),
$h$ is given by
$h = \mathrm{diag}\,\left(h_+, h_-\right)$
where
$h_\pm \equiv -i( \tau_x \partial_x \pm  \tau_y \partial_y)$,
with the Fermi velocity set to one for convenience;
the term $\Delta \Lambda$ with $\Delta \in \mathbb{C}$  and $
\Lambda = \sigma_x \tau_z$,
pairs electrons associated with $\mathbf{K}_+$ and $\mathbf{K}_-$.

Now we examine the core structure of a superconducting vortex in the SC graphene.
The vortex is created by applying magnetic field perpendicular to the graphene plane,
which induces vortex lines in the bulk SC and pass through the graphene layer.
With a single vortex (with vorticity $n$)
centered at $\mathbf{r}=0$ the $\Delta$ in Eq.\ (\ref{schamilton})
becomes $\Delta(|\textbf{r}|) e^{i n\theta}$ where
$\Delta(|\textbf{r}|) \propto |\textbf{r}|^n$ as $|\textbf{r}| \to 0$ and $\Delta(|\textbf{r}|)\to \Delta$ as $|\textbf{r}|\to \infty $. In addition
the magnetic field is introduced via the vector potential through minimal coupling.
In the presence of the vortex the BdG equation
$
K \Psi(\mathbf{r}) = E \Psi(\mathbf{r})
$
acquires two zero energy solutions
(for simplicity we consider single winding vortex)\cite{pgw}.
In the symmetric gauge $\textbf{A}(\textbf{r})=A(|\textbf{r}|)\ \textbf{e}_\theta$,
the zero energy eigenfunctions $\Psi_1(\mathbf{r})$
and
$\Psi_2(\mathbf{r})$
are given by:
\begin{eqnarray}\label{zmodes}
\Psi^{T}_{1}(\textbf{r})
\!\!&=&\!\!
\left[f(|\textbf{r}|),0,0,0,
0,0,0,-i f(|\textbf{r}|)\right],
\nonumber \\%%%%%
\Psi^{T}_{2}(\textbf{r})
\!\!&=&\!\!
\left[0,0,0,-if(|\textbf{r}|),
f(|\textbf{r}|),0,0,0\right],
\end{eqnarray}
where the localized function $f(|\textbf{r}|)$ is given by
$f(|\textbf{r}|)=e^{q \int^{|\mathbf{r}|}_0 A(s) ds} g(|\textbf{r}|)$ with $g(|\textbf{r}|) \to \mbox{\textit{constant}}$
as $|\textbf{r}|\to 0$ and $g(|\textbf{r}|) \to\  e^{-\Delta |\textbf{r}|}$ as
$|\textbf{r}|\to \infty$ ($q>0$ is the electric charge).

Interestingly these degenerate vortex core states form the invariant space of a valley spin  SU(2)
symmetry associated with the valley degeneracy of the BdG Hamiltonian. The three generators of this
SU(2) are
\begin{eqnarray}\label{Qdef}
Q_1= \tau_x\sigma_y\rho_0,
\quad
Q_2= - \tau_x\sigma_x\rho_z,
\quad
Q_3=\tau_0\sigma_z\rho_z,
\end{eqnarray}
with the expectation value of $Q_3$ equal to the difference of the
electron number associated with the two different valleys ${\bf K}_\pm$. It is easy to check that
 $\left[
Q_i, Q_j
\right]
= 2 i \epsilon_{ijk} Q_j
$.
$Q_3$ is diagonal in the representation given by Eq.\ (\ref{zmodes})
, i.e. the two zero modes carry $\pm 1$ $Q_3$ quantum numbers.
In the presence of valley spin SU(2) breaking perturbation the zero-mode degeneracy will be lifted.
Within the subspace spanned by the zero modes such effect is described by
an effective magnetic field, $\boldsymbol{B}_{\mathrm{eff}} \cdot \boldsymbol{\Gamma}$,
where three components of $\boldsymbol{\Gamma}$ are the valley spin Pauli matrices.
With  $\boldsymbol{B}_{\mathrm{eff}}\neq \textbf{0}$
the SU(2) symmetry is broken down to U(1) and we expect the split zero modes to be eigenstates of  the generator of the residual U(1).

The above SC vortex core physics is nicely captured by the following double Chern-Simons action
\begin{eqnarray}
S_{\mathrm{CS}}[A,A^5,\Theta]
=
\frac{1}{2\pi}
\int dt d\mathbf{r}
 \epsilon^{\mu\nu\rho}
A^5_{\mu}  \partial^{\ }_{\nu}
\left(
\partial^{\ }_{\rho}\Theta
+ 2 q A^{\ }_{\rho}
\right),
\label{double CS}
\end{eqnarray}
where $A$ is the ordinary (external) electromagnetic gauge field, $\Theta$ is the phase of the SC order parameter, and $A_5$ is the gauge field that minimally couples to the residual U(1) phase of the valley spin.
By taking the derivative with respect to
the temporal component of $A^5_{\mu}$ in
Eq.\ (\ref{double CS}),
we obtain
$
\Delta Q^5=\delta S_{\mathrm{CS}}/\delta A^5_{0}|_{A^5=0} =
(2\pi)^{-1}
\int d\mathbf{r}
\epsilon^{ij}
\partial_{i} \partial_{j}\Theta
=
n
$,
where $n$ is the vorticity, and $Q^5$ is the generator of the residual U(1) valley spin symmetry.
Formally, the action (\ref{double CS}) can be obtained by
integrating out fermions in the presence of arbitrary background
of $A, A^5$ and $\Theta$
(see Ref.\ \cite{Goldstone1981}).
The double Chern-Simon term (\ref{double CS}) is the formal declaration of the quantum valley Hall effect discussed in the introduction.

With valley spin SU(2) symmetry breaking the SC vortex core acquires an interesting texture
which allows its $\mathrm{SU}(2)$ structure to be identified by experiments.
As we shall discuss below, this texture is in the form of
site or bond charge density wave (CDW).  The site CDW breaks the equivalence between the $A$ and $B$ sublattice without enlarging the unit cell. It is described by a real order parameter. On the other hand the bond density wave (or Kekule order\cite{Hou07}) triples the unit cell size,
and is described by two real order parameters
each corresponding to one of the two complementary Kekule distortion patterns\cite{Hou07}.
In the presence of these
three types of order, the  following ``mass'' terms
$
\int d\textbf{r}
\Psi^{\dag}
M_\alpha
\Psi
$
are added to the Dirac Hamiltonian, where
$
M_{\mathrm{CDW}}\equiv M_3 = \sigma_0 \tau_z \rho_z
$, and $
M_{\mathrm{Kek}1}\equiv M_1=\sigma_1 \tau_2 \rho_0,
$
$
M_{\mathrm{Kek}2}\equiv M_2= \sigma_2 \tau_2 \rho_3 .
$

It turns out that each of the $Q_{1,2,3}$ generates
a rotation of a pair of the density wave order parameters
discussed above,
just like the electric charge operator $Q$ generates
a rotation between the
real and imaginary part  $M_{\mathrm{Re}\, \mathrm{SC}} = \sigma_x \tau_z \rho_x$,  $M_{\mathrm{Im}\, \mathrm{SC}} =\sigma_x \tau_z \rho_y$ of the SC order parameter (see Table \ref{bil}).
\begin{table}[tp]
\begin{ruledtabular}
\begin{tabular}{cc|cc}
&valley spin operator & Rotated pair  & \\   \hline
&    $Q_3$ & $(M_{\mathrm{Kek}1},M_{\mathrm{Kek}1})$ & \\ \hline
& $Q_1$   &   $(M_{\mathrm{CDW}},M_{\mathrm{kek}2})$ & \\ \hline
& $Q_2$   &   $(M_{\mathrm{CDW}},M_{\mathrm{kek}1})$ & \\ \hline
& $Q$ & $(M_{\mathrm{Re}\, \mathrm{SC}},M_{\mathrm{Im}\, \mathrm{SC}})$ & \\
        \end{tabular}
\end{ruledtabular}
\caption{
Valley pseudospin and electric charge generators (left column)
with corresponding order parameters (Higgs mass terms)
which are rotated by the generator
(right column).} \label{bil}\end{table}

One can infer these relations by noting that
under a uniform rotation around $i$-th axis $Q_i$ in the valley spin  space
$
\Psi \to e^{ {i}\varphi_i Q_{i}} \Psi,
$
the Hamiltonian in the presence of
Higgs mass terms $M_{j,k}$
[where $(i,j,k)$ is a cyclic permutation of $(1,2,3)$]
(i.e. $
\mathcal{H}+
\int d\textbf{r} \Psi^\dagger
\left(
n_j M_j +n_k M_k
\right)
\Psi
$)
is invariant if
the rotation in $\Psi$
is accompanied by a simultaneous rotation in $n_{j,k}$,
$
n_{j}+ {i} n_{k}
\to
e^{-2 {i}\varphi_i}
\left(
n_{j}+ {i} n_{k}
\right),
$
as one can check from the commutation relations,
$\left[ Q_{i}, M_{j} \right] = 2 {i} \epsilon_{ijk} M_{k} $.
In turn, this implies the presence of valley spin SU(2) breaking perturbation induces
a linear combination of the three types of order discussed above.
Within the subspace spanned by the vortex zero modes,
it turns out that $B_{\mathrm{eff},1,2,3}\equiv n_{1,2,3}$.

\begin{figure}[htp] \label{CDW} \centering
\begin{center}
\vspace{0.1in}\subfigure[]{\label{cdw1}
\includegraphics[width=0.2 \textwidth]{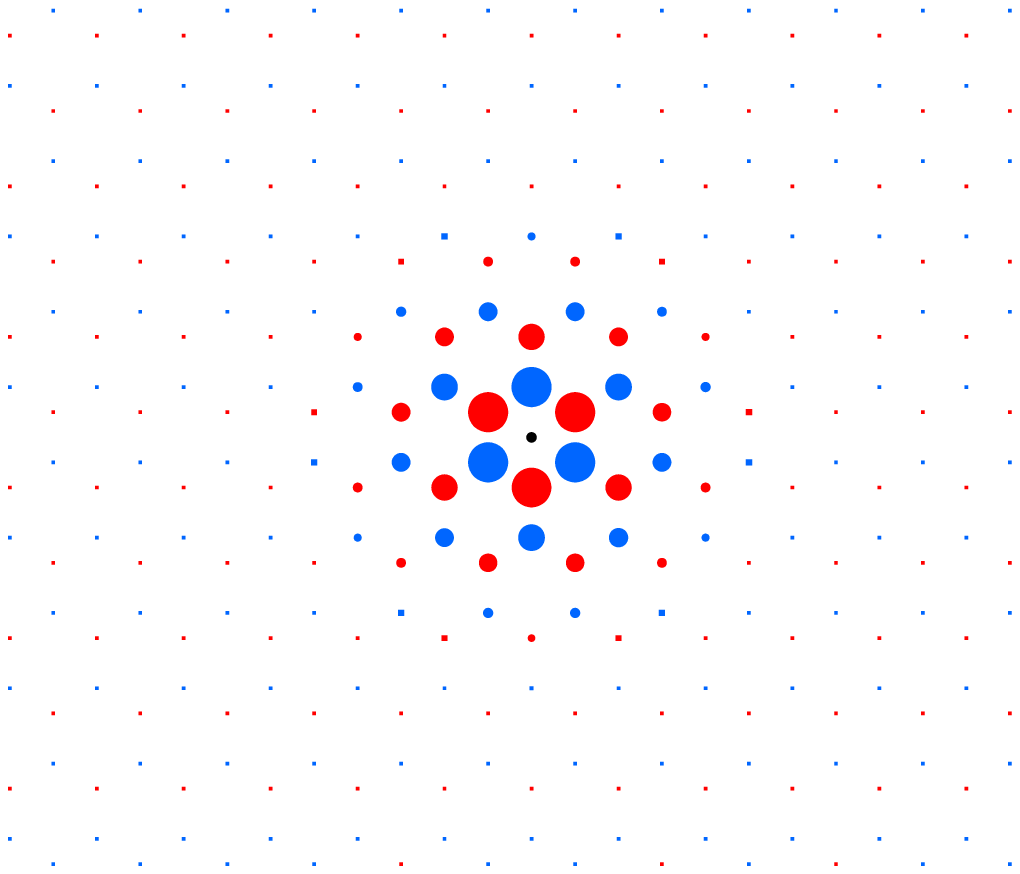} \vspace{0.1in}
} \subfigure[]{
\includegraphics[width=.2 \textwidth]{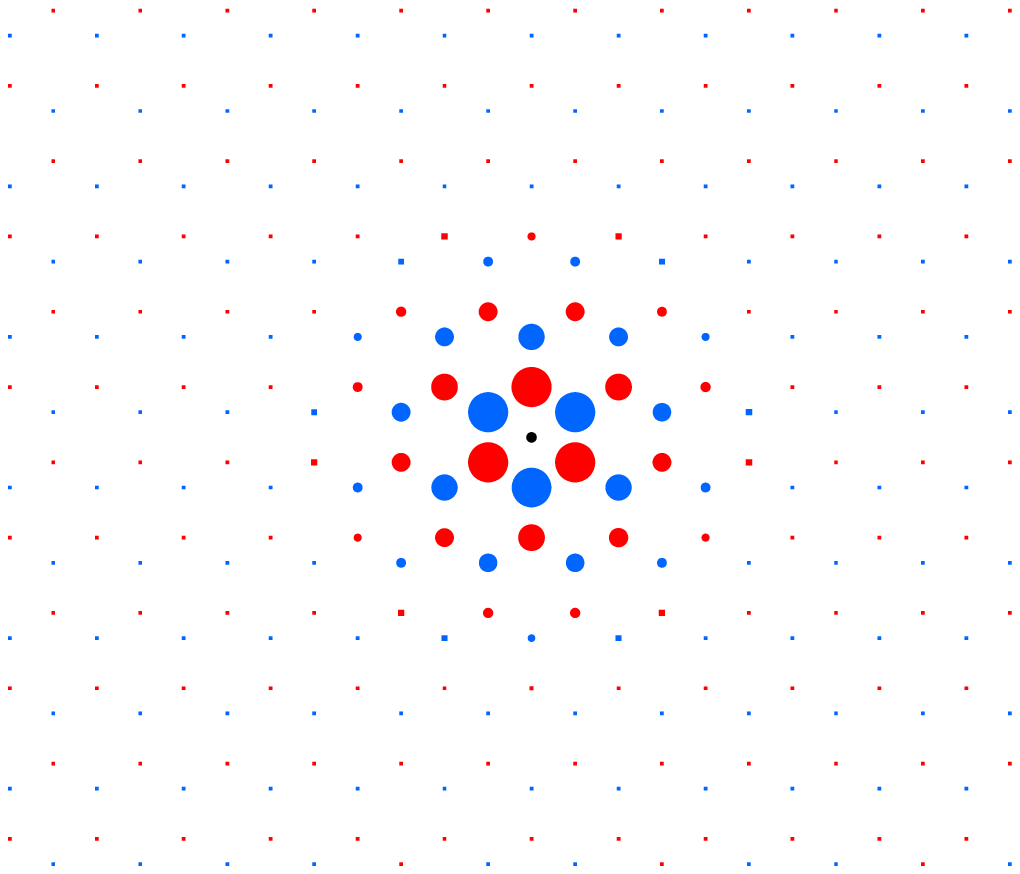}
\vspace{0.1in} \label{cdw2}}
 \caption{
CDW texture of the vortex core states
for (a) positive and (b) negative energy
in the presence of a small staggered chemical potential
around the core.
Blue (red) points show surplus (deficit) quasiparticle density,
with their size proportional to the magnitude.}
\label{sc}\end{center}
\end{figure}

We have studied two different types of valley SU(2) breaking by solving the BdG Hamiltonian numerically on a finite graphene lattice.  In the first case, the interaction of the vortex with the edge of the system induces a
$\boldsymbol{B}_{\mathrm{eff}}$ pointing in the $x$-$y$ plane
leading to a Kekule type vortex texture.
In Fig.\ \ref{sc} we plot the expectation value of the nearest neighbor
link operator
$c^{\dag}_{i s} c^{\ }_{j s}$ in the core states
($c^{\dag}_{i s}$ is the lattice electron operator at $i$-th site with spin $s$).
Here blue (red) denote positive (negative) values and
the thickness represent the amplitude.
It is interesting to note that the Kekule texture associated with the two split core states are opposite.
The second type of valley SU(2) breaking is triggered by explicitly turning on a  staggered chemical potential.
This can be achieved through the interaction with the ``buffer layer'' when graphene
is grown on a substrate\cite{epitax} (Fig.\ \ref{setup}-(b)).
In \textcolor{blue}{Figs.}\ \ref{cdw1} and \ref{cdw2}
we show the induced site CDW texture associated with the two core states.
Similar to the Kekule texture discussed above,
the site CDW texture associated with the two core states are opposite.
In principle these vortex core texture can be probed by scanning tunneling microscope (STM).
In addition to STM the vortex core texture induced by the staggered chemical potential can be detected via a measurement of the orbital magnetic moment.
This is because in this case the
two core states carry
opposite $Q_3$ and it has been shown that in the presence of staggered potential
the graphene quasiparticles carry a finite magnetic orbital
moment which has opposite direction for the two valleys\cite{orbital}.
The complimentary nature of the texture of the two modes is important for differentiating
the effect we are studying from any other such effect generated simply by  translation symmetry breaking.
Also notice that changing the direction of magnetic field in the vortex core
simply change the pattern to its complimentary form which is another unique feature of our result.

\begin{figure}[tp]
\begin{center}
\begin{tabular}{ll}
\begin{minipage}{0.5\hsize}
\includegraphics[width=1 \textwidth]{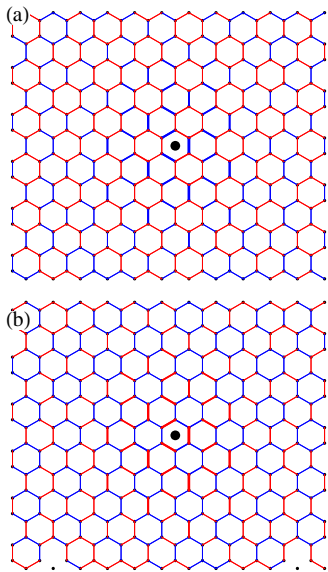}
\end{minipage}
\end{tabular}
 \caption{
Kekule texture observed
for the split core states at
(a) positive and (b) negative energy.
Blue (red) bonds show
strengthened (weakened) bonds, respectively.
The filled circle denotes the position of the vortex.
\label{sc}}
\end{center}
\end{figure}

\begin{figure}[tp]
\begin{center}
\includegraphics[width=0.25 \textwidth]{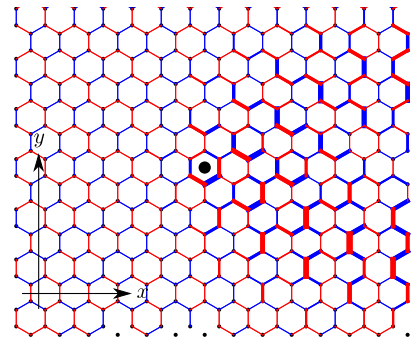}%\nolabel
 \caption{
The Kekule texture in the presence of supercurrent in $y$-direction.
Blue (red) bonds show
strengthened (weakened) bonds, respectively.
The filled circle denotes the position of the vortex.
\label{qvhe}}
\end{center}
\end{figure}

In Fig.\  \ref{qvhe}, we demonstrate the QVHE predicted by Eq.\ (\ref{double CS}).
A  single vortex is created at the center of the central SC graphene region and a finite supercurrent is
 passed along the strip by spatially twisting the phase of SC order parameter.
Due to the interaction with the edges, the vortex texture is of the Kekule type.
The asymmetric Kekule texture can be understood as the $Q^5$ accumulation induced by
ramping up the supercurrent.
This effect is a manifestation of the QVHE when one differentiates
Eq.\ (\ref{double CS}) with respect to $A^5_j$.

Although the accumulation of valley spin by supercurrent
can be viewed as a close relative of the QSHE,
unlike the spin-filtered gapless mode at the QSH edge,
there is no stable edge mode in the QVHE in general.
In the QSH edge,
time-reversal symmetry acting on spin
protects the gapless nature of the edge modes,
i.e. in order to make the edge modes gapped, one needs to break time-reversal
symmetry, say, by adding magnetic impurities or inducing magnetic order at the edge.
On the other hand,
there is no such discrete symmetry acting on valley spin in the QVHE.
In other word, quite generally,
the edge mode of the QVHE is gapped
because of valley spin symmetry breaking,
leading to the Kekule or CDW texture at the edge.

The connection between the physics discussed so far and the Landau forbidden transition discussed in
Ref.\ \cite{dqc} is the following.
Imagine
one starts with the superconducting graphene
and allows the vortices to proliferate.
In the presence of proper symmetry breaking, which, say,
favors the Kekule texture in the vortex core, such proliferation will drive the SC into an insulator with Kekule order. Since the symmetry groups of the two
phases do not have the subgroup relation, the continuous transition triggered by the vortex proliferation
defies the Landau rule.  Conversely, if somehow the Kekule order is stabilized in graphene,
as shown in Ref.\ \cite{Hou07},
the vortex of the Kekule order can carry charge.
As a result when the Kekule vortices proliferate and melt the Kekule order,
a superconducting state emerges.
The same scenario has been proposed originally
for the putative direct transition
between VBS and Neel orders
\cite{dqc,Tanaka-Hu05}.
The duality between SC and VBS (Kekule) orders discussed in
this paper can be mapped to that between the
VBS and Neel orders upon a particle-hole transformation on one spin species.
%$c^{\dag}_{i\downarrow} \to (-1)^i c^{\ }_{i\downarrow}$.
Other types of order parameters,
which are seemingly different, and are not symmetry-related
to each other in the Landau-Ginzberg sense,
can be related in the same way;
the quantum spin Hall phase (Kane-Mele phase)
and $s$-wave SC phase \cite{tarun},
Neel and $d$-wave SC phases \cite{ranvl},
etc.

The above duality is the physics of the Wess-Zumino-Witten (WZW) term.  When several nearly degenerate order parameters (or Higgs fields) compete to open mass gaps in a massless Dirac theory, one can derive a non-linear sigma model to describe the low energy/long wavelength  order parameter fluctuations by integrating out the fermion fields. When the ``mass term'' associated with these orders obey certain anticommutation relation, a purely topological term, the  WZW term, arises. For example, the five types of order parameter we have encountered (real and imaginary parts of superconductivity,
two Kekule order parameters,
and the CDW order) give rise to 5 mass-generating fermion bilinear
$
\int d\textbf{r}\sum_{a=1}^5
\Psi^{\dag}
\phi_a M_a
\Psi
$
in graphene. The five hermitian mass matrices $M_a$ anticommute with each other,
\begin{eqnarray}
\{ M_a, M_b\} = 2 \delta_{ab},
\quad a,b=1,\ldots, 5,
\end{eqnarray}
and also with the kinetic term, $\sigma_0 \tau_x \rho_0p_x+\sigma_z \tau_y \rho_zp_y$, of graphene. 
The WZW term generated upon integrating out the fermion fields\cite{AbanovWiegmann2000} is 
%In a theory where these five order parameters all appear on equal footing in the low energy
%theory, one can view the order parameters as an $\mathrm{O}(5)$
%vector satisfying $\sum_a\phi_a^2=M_0^2.$  In this case one can write
%$\phi_a=M_0n_a$ and integrate out fermions to obtain
%an effective non-linear sigma model
%via computing
%$
%\int \mathcal{D}
%\left[\Psi^{\dag},\Psi\right]
%\exp
%\{
%-S[\Psi^{\dag},\Psi, n]
%\}
%=
%\exp
%\{
%-S_{\mathrm{eff}}[n]
%\}
%$.
%For a slowly varying $n_a$, the fermions spectrum is gapped,
%and they can be integrated out safely at low energy and long wavelength.
%Aside from the usual stiffness
%terms, the resulting non-linear sigma model contains a  WZW term\cite{AbanovWiegmann2000}:
\begin{eqnarray}
&&S_{\mathrm{WZW}}\nonumber \\
&&=
\frac{-2\pi {i}} {\mathrm{Area}(S^4)}
\int^1_0 du \int d^3 x\,
%\nonumber \\%%%%%
%&&
%\quad
%\times
\epsilon^{abcde}
\phi_a \partial_u \phi_b \partial_{\tau} \phi_c \partial_x \phi_d \partial_y \phi_e.\nonumber
\end{eqnarray}
As discussed by Abanov and Wiegmann\cite{AbanovWiegmann2000} this term encodes the
information on soliton charge, duality, etc. What is perhaps less appreciated is the fact that 
the physics of the WZW term is manifested even when the order parameters in question are 
not low energy degrees of freedom. Indeed, in the present paper, the SC order is artificially introduced  by the proximity effect, and the Kekule/CDW order is not closed being realized. Despite that the SC vortex core exhibits the texture of the dual order parameter.  
For graphene there are 36 possible mass terms
which anticommute with the graphene kinetic term,
each representing a gapped order of some kind.
Out of these 36 mass terms,
one can form 56 sets of five mutually anticommuting
mass terms\cite{shinsei}.
Among these, SC-Kekule duality in this paper seems to be
the easiest to probe experimentally.

%Superconductivity in graphene has been a subject of wide investigations\cite{neto}.
%As mentioned before proximity effect induced supercurrent has been reported,
%but whether graphene can spontaneously become a superconductor is still an open question.
%Along with experimental evidence for strong coupling of Kekule phonon modes
%with electrons\cite{kekule,phonon},
%our results also suggest the close connection between Kekule structure and superconductivity in graphene.

{\it Acknowledgements}. DHL was supported by DOE grant number DE-AC02-05CH11231. P.G. acknowledges support from LBNL DOE-504108. S.R. thanks the Center for Condensed Matter Theory at University of California, Berkeley for its support.

\bibliography{grapheneorder}

\begin{thebibliography}{19}
\expandafter\ifx\csname natexlab\endcsname\relax\def\natexlab#1{#1}\fi
\expandafter\ifx\csname bibnamefont\endcsname\relax
  \def\bibnamefont#1{#1}\fi
\expandafter\ifx\csname bibfnamefont\endcsname\relax
  \def\bibfnamefont#1{#1}\fi
\expandafter\ifx\csname citenamefont\endcsname\relax
  \def\citenamefont#1{#1}\fi
\expandafter\ifx\csname url\endcsname\relax
  \def\url#1{\texttt{#1}}\fi
\expandafter\ifx\csname urlprefix\endcsname\relax\def\urlprefix{URL }\fi
\providecommand{\bibinfo}[2]{#2}
\providecommand{\eprint}[2][]{\url{#2}}

\bibitem[{\citenamefont{Novoselov et~al.}(2004)}]{graphene}
\bibinfo{author}{\bibfnamefont{K.~S.} \bibnamefont{Novoselov}}
  \bibnamefont{et~al.}, \bibinfo{journal}{Science}
  \textbf{\bibinfo{volume}{306}}, \bibinfo{pages}{666} (\bibinfo{year}{2004}).

\bibitem[{\citenamefont{Novoselov et~al.}(2005)}]{graphenedirac}
\bibinfo{author}{\bibfnamefont{K.~S.} \bibnamefont{Novoselov}}
  \bibnamefont{et~al.}, \bibinfo{journal}{Nature}
  \textbf{\bibinfo{volume}{438}}, \bibinfo{pages}{197} (\bibinfo{year}{2005}).

\bibitem[{\citenamefont{Heersche et~al.}(2007)}]{graphenesc1}
\bibinfo{author}{\bibfnamefont{H.~B.} \bibnamefont{Heersche}}
  \bibnamefont{et~al.}, \bibinfo{journal}{Nature}
  \textbf{\bibinfo{volume}{446}}, \bibinfo{pages}{56} (\bibinfo{year}{2007}).

\bibitem[{\citenamefont{Shailos et~al.}(2007)}]{graphenesc2}
\bibinfo{author}{\bibfnamefont{A.}~\bibnamefont{Shailos}} \bibnamefont{et~al.},
  \bibinfo{journal}{Euro. Phys. Lett.} \textbf{\bibinfo{volume}{79}},
  \bibinfo{pages}{57008} (\bibinfo{year}{2007}).

\bibitem[{\citenamefont{Beenakker}(2006)}]{beenakker}
\bibinfo{author}{\bibfnamefont{C.~J.} \bibnamefont{Beenakker}},
  \bibinfo{journal}{Phys. Rev. Lett.} \textbf{\bibinfo{volume}{97}},
  \bibinfo{pages}{067007} (\bibinfo{year}{2006}).

\bibitem[{\citenamefont{Kane and Mele}(2006)}]{qshkm}
\bibinfo{author}{\bibfnamefont{C.}~\bibnamefont{Kane}} \bibnamefont{and}
  \bibinfo{author}{\bibfnamefont{E.}~\bibnamefont{Mele}},
  \bibinfo{journal}{Phys. Rev. Lett.} \textbf{\bibinfo{volume}{95}},
  \bibinfo{pages}{226801} (\bibinfo{year}{2006}).

\bibitem[{\citenamefont{Bernevig et~al.}(2006)}]{qshe}
\bibinfo{author}{\bibfnamefont{B.}~\bibnamefont{Bernevig}}
  \bibnamefont{et~al.}, \bibinfo{journal}{Science}
  \textbf{\bibinfo{volume}{314}}, \bibinfo{pages}{1757} (\bibinfo{year}{2006}).

\bibitem[{\citenamefont{Senthil et~al.}(2004)}]{dqc}
\bibinfo{author}{\bibfnamefont{T.}~\bibnamefont{Senthil}} \bibnamefont{et~al.},
  \bibinfo{journal}{Science} \textbf{\bibinfo{volume}{303}},
  \bibinfo{pages}{1490} (\bibinfo{year}{2004}).

\bibitem[{\citenamefont{Hou et~al.}(2007)\citenamefont{Hou, Chamon, and
  Mudry}}]{Hou07}
\bibinfo{author}{\bibfnamefont{C.-Y.} \bibnamefont{Hou}},
  \bibinfo{author}{\bibfnamefont{C.}~\bibnamefont{Chamon}}, \bibnamefont{and}
  \bibinfo{author}{\bibfnamefont{C.}~\bibnamefont{Mudry}},
  \bibinfo{journal}{Phys. Rev. Lett.} \textbf{\bibinfo{volume}{98}},
  \bibinfo{pages}{186809} (\bibinfo{year}{2007}).

\bibitem[{\citenamefont{Demler et~al.}(2001)}]{sachdev}
\bibinfo{author}{\bibfnamefont{E.}~\bibnamefont{Demler}} \bibnamefont{et~al.},
  \bibinfo{journal}{Phys. Rev. Lett.} \textbf{\bibinfo{volume}{87}},
  \bibinfo{pages}{067202} (\bibinfo{year}{2001}).

\bibitem[{\citenamefont{Ghaemi and Wilczek}(2007)}]{pgw}
\bibinfo{author}{\bibfnamefont{P.}~\bibnamefont{Ghaemi}} \bibnamefont{and}
  \bibinfo{author}{\bibfnamefont{F.}~\bibnamefont{Wilczek}}
  (\bibinfo{year}{2007}), \eprint{arXiv:0709.2626}.

\bibitem[{\citenamefont{Goldstone and Wilczek}(1981)}]{Goldstone1981}
\bibinfo{author}{\bibfnamefont{J.}~\bibnamefont{Goldstone}} \bibnamefont{and}
  \bibinfo{author}{\bibfnamefont{F.}~\bibnamefont{Wilczek}},
  \bibinfo{journal}{Phys. Rev. Lett.} \textbf{\bibinfo{volume}{47}},
  \bibinfo{pages}{986} (\bibinfo{year}{1981}).

\bibitem[{\citenamefont{Zhou et~al.}(2007)}]{epitax}
\bibinfo{author}{\bibfnamefont{S.~Y.} \bibnamefont{Zhou}} \bibnamefont{et~al.},
  \bibinfo{journal}{Nature Mat.} \textbf{\bibinfo{volume}{6}},
  \bibinfo{pages}{770} (\bibinfo{year}{2007}).

\bibitem[{\citenamefont{Xia et~al.}(2007)}]{orbital}
\bibinfo{author}{\bibfnamefont{D.}~\bibnamefont{Xia}} \bibnamefont{et~al.},
  \bibinfo{journal}{Phys. Rev. Lett.} \textbf{\bibinfo{volume}{99}},
  \bibinfo{pages}{236809} (\bibinfo{year}{2007}).

\bibitem[{\citenamefont{Tanaka and Hu}(2005)}]{Tanaka-Hu05}
\bibinfo{author}{\bibfnamefont{A.}~\bibnamefont{Tanaka}} \bibnamefont{and}
  \bibinfo{author}{\bibfnamefont{X.}~\bibnamefont{Hu}}, \bibinfo{journal}{Phys.
  Rev. Lett.} \textbf{\bibinfo{volume}{95}}, \bibinfo{pages}{036402}
  (\bibinfo{year}{2005}).

\bibitem[{\citenamefont{Grover and Senthil}(2008)}]{tarun}
\bibinfo{author}{\bibfnamefont{T.}~\bibnamefont{Grover}} \bibnamefont{and}
  \bibinfo{author}{\bibfnamefont{T.}~\bibnamefont{Senthil}}
  (\bibinfo{year}{2008}), \eprint{arXiv:0801.2130}.

\bibitem[{\citenamefont{Ran et~al.}(2008)}]{ranvl}
\bibinfo{author}{\bibfnamefont{Y.}~\bibnamefont{Ran}} \bibnamefont{et~al.}
  (\bibinfo{year}{2008}), \eprint{arXiv:0806.2321}.

\bibitem[{\citenamefont{Abanov and Wiegmann}(2000)}]{AbanovWiegmann2000}
\bibinfo{author}{\bibfnamefont{A.~G.} \bibnamefont{Abanov}} \bibnamefont{and}
  \bibinfo{author}{\bibfnamefont{P.~B.} \bibnamefont{Wiegmann}},
  \bibinfo{journal}{Nucl. Phys. B} \textbf{\bibinfo{volume}{570}},
  \bibinfo{pages}{685} (\bibinfo{year}{2000}).

\bibitem[{\citenamefont{Ryu et~al.}()}]{shinsei}
\bibinfo{author}{\bibfnamefont{S.}~\bibnamefont{Ryu}} \bibnamefont{et~al.},
  \eprint{in preparation}.

\end{thebibliography}

\end{document}